\documentclass[envcountsame]{llncs}
\pdfoutput=1 
\usepackage[utf8]{inputenc}
\usepackage[T1]{fontenc}
\usepackage[english]{babel} 

\usepackage{etex}
\usepackage{xargs,ifthen}
\usepackage{fixltx2e} 			
\usepackage{lmodern}				

\usepackage{algorithm,algorithmic}

\usepackage{tikz}
\usetikzlibrary{automata,positioning}
\usetikzlibrary{arrows,calc}

\usepackage{blindtext}
\usepackage{microtype}
\usepackage{csquotes}
\usepackage{enumitem}
	\setitemize{itemsep=1pt,topsep=3pt,parsep=0pt,partopsep=0pt}
\usepackage{xcolor}
\definecolor{persimmon}{rgb}{0.93, 0.35, 0.0}
\definecolor{cssgreen}{rgb}{0.0, 0.5, 0.0}
\usepackage{hyperref}
	\hypersetup{
		plainpages			=	false,		
		bookmarksnumbered	=	true,		
		bookmarksopen		=	true,		
		bookmarksdepth		=	2,			
		colorlinks			=	true, 		
		breaklinks			=	true,		
		linkcolor			=	persimmon,	
		citecolor			=	cssgreen,	
		pdfdisplaydoctitle	= 	true,		
	}
\usepackage{xspace}
\usepackage{multicol}
\usepackage{wrapfig}
\usepackage{graphicx}

\usepackage{amsmath,amssymb}
\usepackage{mathtools}

\usepackage{stmaryrd}
\usepackage{cmll}

\addto\extrasenglish{%

\renewcommand{\subsectionautorefname}{Sect.}
}

\newcommand{\aref}[1]{\hyperref[#1]{Appendix~\ref{#1}}} 

\usepackage{dialogue}

\spnewtheorem*{notation}{Notation}{\itshape}{\rmfamily}
\spnewtheorem*{notations}{Notations}{\itshape}{\rmfamily}

\newcommand{\safespace}[1]		
{
	
	\vspace{#1}
	
}

\newcommand{\C}[1]{\mathcal{#1}}

\newcommand{\BF}[1]{\mathbf{#1}}

\newcommand{\IT}[1]{\mathit{#1}}
\newcommand{\TT}[1]{\mathtt{#1}}

\newcommand{\flow}{\ensuremath{\leftharpoonup}}		
\newcommand{\sflow}{\ensuremath{\leftrightharpoons}}		

\newcommand{\fact}[1]{#1\flow\factsym}							
\newcommand{\wiring}[1]{#1}										
\newcommand{\mgu}[1]{\IT #1}										
\newcommand{\obs}[1]{\ensuremath{#1}}							

\newcommand{\balg}{{\C U_b}}	
\newcommand{\dalg}{\C U_d}	
\newcommand{\ualg}{\C U}	
\newcommand{\calg}[1]{#1^\leftharpoonup}	
\newcommand{\walg}{\C W}		
\newcommand{\ralg}[2]{#2^{\setminus #1}} 
\newcommand{\oalg}{\C O} 
\newcommand{\matr}[2]{\C M_{#1}\left(#2\right)}

\newcommand{\ptensor}{\,\dot{\otimes}\,}		

\newcommand{\comp}[1]{\BF{Comp}(#1)}
\newcommand{\G}{\BF G}
\newcommand{\h}{\TT h} 
\newcommand{\M}{\TT M}
\newcommand{\A}{\TT A}

\renewcommand{\L}{\TT{L}}
\newcommand{\R}{\TT{R}} 
\newcommand{\LR}{\TT{LR}}
\newcommand{\start}{\star}
\newcommand{\p}{\scalebox{0.6}{$\,\bullet\,$}}

\newcommand{\factsym}{\star}
\newcommand{\nary}[1]{\TT A_{#1}}

\newcommand{\wordterm}[5]{\TT{#1} \p #2 \p #3 \p #4 \p \M(#5)}

\newcommand{\setofrep}[1]{\C{R}(#1)} 

\newcommand{\isnilp}[1]{\BF{Nil}(#1)} 

\newcommand{\extlist}[2]{#1,\,\dots\,,#2}		
\newcommand{\extset}[2]{\left\{\,#1,\,\dots\,,#2\,\right\}}		
\newcommand{\set}[2]{\left\{\:#1 \ \middle|\ #2\:\right\}}	
\newcommand{\unitalg}{\C I}					
\newcommand{\unit}{I}						
\newcommand{\numberset}[1]{\ensuremath{\mathbb{#1}}}	
\renewcommand{\natural}{\numberset{N}} 					
\newcommand{\naturalp}{\numberset{N^*}} 				
\renewcommand{\max}{\mathop{\mathrm{max}}}

\newcommand{\proglang}[1]{\textsc{#1}}		
\newcommand{\var}{\TT{Var}}									
\newcommand{\vars}{\TT{V}}				
\newcommand{\terms}{\TT{T}}				
\newcommand{\langage}[1]{\mathcal{#1}}	
\newcommand{\lang}[1]{\langage{L}(#1)}	
\newcommand{\state}[1]{\BF{#1}}

\newcommand{\GoI}{\proglang{GoI}\xspace}
\newcommand{\systF}{System~\proglang{F}\xspace}
\newcommand{\LJs}{\proglang{LJ$^2$}\xspace}

\newcommand{\card}[1]{\mathop{\mathrm{card}}(#1)}
\newcommand{\cc}[1]{\textsc{#1}\xspace}					
\newcommand{\logspace}{\cc{D\-Log\-space}}							
\newcommand{\nlogspace}{\cc{N\-Log\-space}}							
\newcommand{\conlogspace}{\cc{co\nlogspace}}		
\newcommand{\ptime}{\cc{Ptime}}								
\newcommand{\nc}{\cc{NC}}								

\newcommand{\ie}{\textit{i.e.}\xspace}
\renewcommand{\iff}{\text{iff}\xspace}

\begin{document}
	\renewcommand{\labelitemi}{\scalebox{0.8}{$\bullet$}} 
	
	\hyphenpenalty=2000
	\tolerance=600


\title{Logic Programming and Logarithmic Space}
\author{
	Clément Aubert\inst{1} \and Marc Bagnol\inst{1} \and Paolo Pistone\inst{1} \and Thomas Seiller\inst{2}
	\thanks{This work was partly supported by the ANR-10-BLAN-0213 Logoi and the ANR-11-BS02-0010 Récré.}}
\institute{Aix Marseille Université, CNRS, 
			I2M UMR 7373, 13453, Marseille, France
			\and I.H.\'{E}.S., Le Bois-Marie, 35, Route de Chartres, 91440 Bures-sur-Yvette, France}

\hypersetup{
	pdfinfo={
		Author		=	{Clément Aubert, Marc Bagnol, Paolo Pistone, Thomas Seiller},
		Title		=	{Logic Programming and Logarithmic Space},
		Keywords	=	{Implicit Complexity, Unification, Logic Programming, Logarithmic Space, Proof Theory, Pointer Machines, Geometry of Interaction, Automata},
		Creator		=	{PDFLaTeX},
		Producer	=	{PDFLaTeX},
		Lang		=	{en}
	}
}

\maketitle

\begin{abstract}
	We present an algebraic view on logic programming, related to proof theory and more specifically linear logic and geometry of interaction.
Within this construction, a characterization of logspace (deterministic and non-deterministic) computation is given \textit{via} a synctactic restriction, using an encoding of words that derives from proof theory.

We show that the acceptance of a word by an observation (the counterpart of a program in the encoding) can be decided within logarithmic space, by reducing this problem to the acyclicity of a graph.
We show moreover that observations are as expressive as two-ways multi-heads finite automata, a kind of pointer machines that is a standard model of logarithmic space computation.

\end{abstract}

\keywords{Implicit Complexity, Unification, Logic Programming, Logarithmic Space, Proof Theory, Pointer Machines, Geometry of Interaction, Automata}

\section{Introduction}
	\vspace{-3mm}
\subsubsection{Proof Theory and Implicit Complexity Theory}
Very generally, the aim of implicit complexity theory (ICC) is to describe complexity classes with no explicit reference to cost bounds: through a type system or a weakened recursion scheme for instance.
The last two decades have seen numerous works relating proof theory
(more specifically linear logic \cite{Girard1987}) and ICC,
the basic idea being to look for restricted substructural logics~\cite{Girard1995} with 
an expressiveness that corresponds exactly to some complexity class.

This has been achieved by various syntactic restrictions, which entail a less complex
(any function provably total in second-order Peano Arithmetic~\cite{Girard1987} can be encoded in second-order linear logic)
cut-elimination procedure: control over the modalities \cite{Schopp2007,Lago2010},
type assignments \cite{Gaboardi2012} or stratification properties \cite{Baillot2010},
to name a few.

\subsubsection{Geometry of Interaction} Over the recent years, the cut-elimination procedure and 
its mathematical modeling has become a central topic in proof theory.
The aim of the geometry of interaction research program~\cite{Girard1989a} is to provide the tools for such a modeling~\cite{Asperti1994,Laurent2001,Seiller2012a}.

As for complexity theory, these models allow for a more synthetic and abstract study of the resources needed to compute the normal form of a program, leading to some complexity characterization results \cite{Baillot2001,Girard2012,Aubert2014}.

\subsubsection{Unification} The unification technique is one of the key-concepts of theoretical computer science: it is a classical subject of study for complexity theory and a tool with a wide range of applications, including logic programming and type inference algorithms.

Unification has also been used to build syntactic models of geometry of interaction~\cite{Girard1995a,Baillot2001,Girard2013} where first-order terms with variables allow for a manipulation of infinite sets through a finite language.

\subsubsection{Logic Programming} After the work of Robinson on the resolution procedure, logic programming has emerged as a new computation paradigm with concrete realizations such as the languages \proglang{Prolog} and \proglang{Datalog}.

On the theoretical side, a lot of efforts has been made to clarify expressiveness 
and complexity issues~\cite{Dantsin2001}:
most problems arising from logic programming are undecidable in their most general form
and some restrictions must be introduced in order to make them tractable.
For instance, the notion of \emph{finitely ground program}~\cite{Calimeri2008} is related to our approach.

\subsubsection{Pointer Machines}
Multi-heads finite automata provide an elegant characterization of logspace computation, in terms of the (qualitative) type of memory used rather than the (quantitative) amount of tape consumed.
Since they can scan but not modify the input, they are usually called \enquote{pointer machines}.

This model was already at the heart of previous works relating geometry of interaction and complexity theory~\cite{Girard2012,Aubert2012,Aubert2014}.

\medskip

\subsubsection{Contribution and Outline}
We begin by exposing the idea of relating geometry of interaction and logic programming, already evoked~\cite{Girard1995a} but never really developed, and by recalling the basic notions on unification theory needed for this article and some related complexity results.

We present in \autoref{sec_ualg} the algebraic tools
used later on to define the encoding of words and pointer machines.%
\renewcommand{\subsectionautorefname}{Section} 
\autoref{sec_balanced}%
\renewcommand{\subsectionautorefname}{Sect.} 
and \autoref{sec_compgraph} introduce the syntactical
 restriction and associated tools that allow us to characterize logarithmic space computation.
Note that, compared to earlier work~\cite{Aubert2014}, 
we consider a much wider class of programs while preserving bounded space evaluation.

The encoding of words enabling our results,
which comes from the classical (Church) encoding of lists in proof theory,
is given in \autoref{sec_representation}.
It allows to define the counterpart of programs, and a notion of acceptance of a word by a program.

Finally, \autoref{sec_complexity} makes use of the tools introduced earlier
to state and prove our complexity results. While the expressiveness part 
is quite similar to earlier presentations~\cite{Aubert2012,Aubert2014}, the proof that acceptance 
can be decided within logarithmic space has been made more modular
by reducing it to cycle search in a graph.

	\subsection{Geometry of Interaction and Logic Programming}\label{ssec_goilp}
	
The geometry of interaction program (\GoI for short), started in 1989~\cite{Girard1989b}, aims at describing the dynamics of computation by developing a fully mathematical model of cut-elimination.
The original motivations of \GoI must be traced back, firstly, to the \emph{Curry-Howard correspondence} between sequent calculus derivations and typed functional programs: it is on the basis of this correspondence that \hbox{cut-elimination} had been proposed by proof-theorists as a paradigm of computation; secondly, to the finer analysis of cut-elimination coming from linear logic~\cite{Girard1987} and the replacement of sequent calculus derivations with simpler geometrical structures (proof-nets), more akin to a purely mathematical description.

In the first formulation of \GoI~\cite{Girard1989a}, derivations in second order intuitionistic logic \LJs (which can be considered, by \emph{Curry-Howard}, as programs in \systF) are interpreted as pairs $(U,\sigma)$ of elements (called \emph{wirings}) of a $\mathbb{C}^{*}$-algebra, $U$ corresponding to the axioms of the derivation and $\sigma$ to the cuts.

The main property of this interpretation is \emph{nilpotency}: there exists an integer $n$ such that $(\sigma U)^{n}=0$.
The cut-elimination (equivalently, the normalization) procedure is then interpreted by the application of an \emph{execution operator}

\medskip
\centerline{$EX(U,\sigma)=\sum_{k}(\sigma U)^{k}$}
\medskip
From the viewpoint of proof theory and computation, nilpotency corresponds to the \emph{strong normalization property}: the termination of the normalization procedure with any strategy.

Several alternative formulations of geometry of interaction have been proposed since 1988 (see for instance~\cite{Asperti1994,Laurent2001,Seiller2012a}); in particular, wirings can be described as logic programs~\cite{Girard1995a,Baillot2001,Girard2013} made of particular clauses called \emph{flows}, which will be defined in \autoref{sec_flows}.


In this setting the resolution rule induces a notion of product of wirings (\autoref{def_wirings}) and in turn a structure of semiring: the \emph{unification semiring} $\ualg$, which can replace the $\mathbb{C}^{*}$-algebras of the first formulations of \GoI\footnote{By adding complex scalar coefficients, one can actually extend $\ualg$ into a $\mathbb{C}^{*}$-algebra~\cite{Girard1995a}.}.

The $EX(.)$ operator of wirings can be understood as a way to compute the fixed point semantics of logic programs. The nilpotency property of wirings means then that the fixed point given by $EX(.)$ is finite, which is close to the notion of \emph{boundedness}~\cite{Dantsin2001} of logic programs.


\smallskip
In definitive, from the strong normalization property for intuitionistic second order logic (or any other system which enjoys a \GoI interpretation), one obtains through the \GoI interpretation a family of bounded (nilpotent) logic programs computing the recursive functions typable in \systF.

This is quite striking in view of the fact that to decide whenever a program is \emph{bounded}%
\footnote{A program is \emph{bounded} if there is an integer $k$ such that the fixed point computation of the program is stable after $k$ iterations, independently of the facts inputed.}
is --~even with drastic constraints~-- an undecidable problem~\cite{Hillebrand1995}, and that in general boundedness is a property that is difficult to ensure.

	\subsection{Unification and Complexity}\label{ssec_unif}
	Unification is a classical subject of research, at the intersection of practical and theoretical considerations.
We recall in the following some notations and some of the numerous links between complexity and unification, and by extension logic programming.


\begin{notations}
	\label{def_terms}
		We consider a set of first-order terms $\terms$, 
		assuming an infinite number of variables $x, y,z,\:\ldots\in\vars$, 
		a binary function symbol $\p$ (written in \emph{infix notation}),
		infinitely many constant symbols $\TT a,\TT b,\TT c, \ldots$
		including the (multipurpose) dummy symbol $\factsym$ and, for any $n \in \naturalp$, at least one $n$-ary function symbol $\nary n$.
	
	Note that the binary function symbol $\p$ is not associative. However, we will write it
by convention as \emph{right associating} to lighten notations: $t\p u\p v := t \p(u\p v)$.
	
		For any $t \in \terms$, we write $\var(t)$ the set of variables occurring in $t$
		(a term is \emph{closed} when $\var(t)=\varnothing$)
		and $\h(t)$ the \emph{heigth} of $t$: the maximal distance from the root to any other subterm in the tree structure of $t$.
		
		The \emph{height of a variable occurrence} in a term $t$ is its distance from the root in the tree structure of the term.\label{def-height}
	\label{renaming}
		A \emph{substitution} $\mgu{\theta}$ is a mapping form variables to terms such that $x \mgu{\theta} = x$ for all but finitely many $x \in \vars$.
		A \emph{renaming} is a substitution $\mgu{\alpha}$ mapping variables to variables and that is bijective.
		A term $t'$ is a \emph{renaming} of $t$ if $t'=t \mgu{\alpha}$ for some renaming $\mgu{\alpha}$.
\end{notations}


\begin{definition}[unification, matching and disjointness]\label{disjoint}
	Two terms $t, u$ are 
	\begin{itemize}
	\item \emph{unifiable} if there exists a substitution%
	\footnote{Called a \emph{unifier}. Remember that two terms $t,u$ that are unifiable have a \emph{most general unifier} (MGU): a unifier $\theta$ such that any other unifier of $t,u$ is an instance of $\theta$.}
	$\mgu{\theta}$ such that $t \mgu{\theta} = u \mgu{\theta}$,
	\item \emph{matchable} if $t',u'$ are unifiable,  where $t',u'$ are renamings of $t,u$ such that $\var(t') \cap \var(u') = \varnothing$,
	\item \emph{disjoint} if they are not matchable.
	\end{itemize}
\end{definition}


It is well-known that the problem of deciding whether two terms are unifiable is $\ptime$-complete \cite[Theorem 1]{Dwork1984}, which implies that parallel algorithms for this problem do not improve much on serial ones.
Finding classes of terms where the MGU research can be efficiently parallelized is a real challenge.

It has been proven that this problem remains $\ptime$-complete even if the arity of the function symbols or the height of the terms is bounded \cite[Theorems 4.2.1 and 4.3.1]{Ohkubo1987}, if both terms are linear or if they do not share variables \cite{Dwork1984,Dwork1988}.
More recently \cite{Bellia2003}, an innovative constraint on variables helped to discover an upper bound of the unification classes that are proven to be in \nc.

Regarding space complexity, the result stating that the \emph{matching problem} is in \logspace~\cite{Dwork1984} (recalled as \autoref{thm-unif-logspace}) will be used in \autoref{sec_soundness}.




\section{The Unification Semiring}\label{sec_ualg}
	This section presents the technical setting of this work, the \emph{unification semiring}: an algebraic structure with a composition law based on unification, that can be seen as an algebraic presentation of a fragment of logic programming.

	\subsection{Flows and Wirings}\label{sec_flows}

Flows can be thought of as very specific Horn clauses: 
safe (the variables of the head must occur in the body) clauses with exactly one atom in the body.

As it is not relevant to this work, we make no difference between predicate symbols and function symbols, for it makes the presentation easier.

\begin{definition}[flows]\label{def_flow}
	A \emph{flow} is a pair of terms $t\flow u$
	with $\var(t)\subseteq\var(u)$.
	Flows are considered up to renaming: for any renaming $\alpha$,
	$t \flow u\,=\,t\alpha \flow u\alpha$.
\end{definition}

Facts, that are usually defined as ground (using only closed terms) clauses with an empty body, can still be represented as a special kind of flows.

\begin{definition}[facts]\label{def_fact}
	A \emph{fact} is a flow of the form $\fact t$.
\end{definition}

\remark{Note that this implies that $t$ is closed.}

\medskip

The main interest of the restriction to flows is that it yields an algebraic structure:
a semigroup with a partially defined product.

\begin{definition}[product of flows]
	Let $u\flow v$ and $t\flow w$ be two flows.
	Suppose we have representatives of the renaming classes
	such that $\var(v)\cap\var(w)=\varnothing$.
	The \emph{product} of $u\flow v$ and $t\flow w$ is defined
	if $v,t$ are unifiable with MGU $\theta$
	as
	$(u\flow v)(t\flow w)\,:=\:u\theta \flow w\theta$.
\end{definition}

\remark{The condition on variables ensures that facts form a \enquote{left ideal} of the set of flows:
if $\BF u$ is a fact and $f$ a flow, then $f\BF u$ is a fact when it is defined.
}

\example{~
\label{example_flows}

$(\TT f(x) \flow x)(\TT f(x)\flow \TT g(x)) = \TT f(\TT f(x)) \flow \TT g(x)$

$(x \p \TT c \flow (y\p y) \p x)(( \TT c \p \TT c) \p x \flow y\p x) = x \p \TT c \flow  \TT c \p \TT x$

$(\TT f(x\p c) \flow x\p\TT d)(\fact{\TT d\p\TT d}) =\fact{\TT f(\TT d\p\TT c)}$
}

\medskip
The product of flows corresponds to the resolution rule in the following sense: given two flows $f=u\flow v$ and $g=t\flow w$ and a $MGU$ $\theta$ of $v$ and $t$, then the resolution rule applied to $f$ and $g$ would yield $fg$.

\smallskip
Wirings then correspond to logic programs (sets of clauses) and the nilpotency condition can be seen as an algebraic variant of the notion of boundedness of these programs.

\begin{definition}[wirings]\label{def_wirings}
	\emph{Wirings} are finite sets of flows.
	The product of wirings is defined as
	$FG:=\set{fg}{f \in F, \, g \in G, \: fg \text{ defined}}$.
	
	We write $\ualg$ the set of wirings and refer to it as the \emph{unification semiring}.
\end{definition}


\medskip
The set of wirings $\ualg$ has a structure of semiring. We use an \emph{additive notation} for sets of flows to stress this point:
\begin{itemize}
	\item The symbol $+$ will be used in place of $\cup$.
	\item We write sets as the sum of their elements: $\extset{f_1}{f_n}:= f_1 + \cdots + f_n$.
	\item We write $0$ for the empty set.
	\item The unit is $\unit := x \flow x$.
\end{itemize}

We will call \emph{semiring} any subset $\C A$ of $\ualg$ such that
\begin{itemize}
	\item $0\in\C A$,
	\item if $\extlist{F_1}{F_n} \in \C A$, then $F_1 + \, \cdots \,+ F_n \in \C A$,
	\item if $F \in \C A$ and $G\in\C A$ then $FG \in \C A$.
\end{itemize}

\begin{definition}[nilpotency]\label{def_nilp}
	A wiring $\wiring{F}$ is \emph{nilpotent} if $\wiring{F}^n=0$ for some $n \in \natural$.
	We may use the notation $\isnilp F$ to express the fact that $F$ is nilpotent.
\end{definition}

As mentionned in \autoref{ssec_goilp}, nilpotency is related with the notion of \emph{boundedness}~\cite{Dantsin2001} of a logic program.
Indeed, if we have a wiring $F$ and a finite set of facts $\BF U$, let us consider the set of facts that can be obtained through $F$ $\set{\BF u}{\BF u\in F^n\BF U \text{ for some }n}$ which can also be written as
%
$(\unit+F+F^2+\cdots)\BF U$ or $EX(F)\BF U$ (where EX(.) is the execution operator of \autoref{ssec_goilp}).

If $F$ is nilpotent, one needs to compute the sum only up to a finite rank that does not depend on $\BF U$, which implies the  boundedness property.



\smallskip
Among wirings, those that can produce at most one fact from any fact will be of interest when considering deterministic \textit{vs.} non-deterministic computation.

\begin{definition}[deterministic wirings]\label{def_fun}
	A wiring $F$ is \emph{deterministic} if given any fact $\BF u$, $\card{F\BF u}\leq 1$.
	We will write $\dalg$ the set of deterministic wirings.
\end{definition}

It is clear from the definition that $\dalg$ forms a semiring.
The lemma below gives us a class of wirings that are deterministic and easy to recognize, due to its more syntactic definition.

\begin{lemma}\label{lem_det}
	Let $F=\sum_i u_i\flow t_i$. If the $t_i$ are pairwise disjoint (\autoref{disjoint}), then $F$ is deterministic.
\end{lemma}

\begin{proof}
	Given a closed term $t$ there is at most one of the $t_i$ that matches $t$, therefore $F(\fact t)$ is either a single fact or $0$.\qed
\end{proof}

	\subsection{The Balanced Semiring}\label{sec_balanced}
	In this section, we study a constraint on variable height of flows which we call \emph{balance}. This syntactic constraint can be compared with similar ones proposed in order to get logic programs that are \emph{finitely ground}~\cite{Calimeri2008}: balanced wirings are a special case of \emph{argument-restricted} programs in the sense of \cite{Lierler2009}.

Balanced wirings will enjoy properties (see \autoref{sec_compgraph}) that will allow to decide their nilpotency efficiently. 

\begin{definition}[balance]%
\label{def-balanced}
	A flow $f=t\flow u$ is \emph{balanced} if for any variable $x\in\var(t)\cup\var(u)$,
	all occurrences of $x$ in either $t$ or $u$ have the same height (recall notations p.~\pageref{def-height})
	which we write $\h_f(x)$, the \emph{height of $x$ in $f$}.
	A wiring $F$ is \emph{balanced} if it is a sum of balanced flows.
	
	We write $\balg$ the set of balanced wirings and refer to it as the \emph{balanced semiring}.
\end{definition}

Note that in \autoref{example_flows}, only the second line shows product of balanced flows.

\begin{definition}[height]\label{def_height}
	The \emph{height} $\h(f)$ of a flow $f=t \flow u$ is $\max\{\h(t),\h(u)\}$.
	The \emph{height} $\h(F)$ of a wiring $F$ is the maximal height of flows in it.
\end{definition}

The following lemma summarizes the properties that are preserved by the product of balanced flows.
It implies in particular that $\balg$ is indeed a semiring.

\begin{lemma}\label{lem_balanced}
	When it is defined, the product $fg$ of two balanced flows $f$ and $g$ is still balanced
	and its height is at most $\max\{\h(f),\h(g)\}$.
\end{lemma}

\begin{proof}[sketch]
	By showing that the variable height condition and the global height are both preserved
	by the basic steps of the unification procedure.
	\qed
\end{proof}


	\subsection{The Computation Graph}\label{sec_compgraph}
	The main tool for a space-efficient treatment of balanced wirings
is an associated notion of graph. 
This section focuses on the algebraic aspects of this notion, 
proving various technical lemmas, and leaves the complexity issues to \autoref{sec_soundness}.

\smallskip

A separating space can be thought of as a finite subset of the Herbrand universe associated with a logic program, containing enough information to decide the problem at hand.

\begin{definition}[separating space]
	A \emph{separating space} for a wiring $F$ is a set of facts $\BF S$ such that
	\begin{itemize}
		\item For all $\BF u\in \BF S$, $F\BF u\subseteq \BF S$.
		\item $F^n\BF u=0$ for all $\BF u\in\BF S$ implies $F^n=0$.
	\end{itemize}
\end{definition}

We can define such a space for balanced wirings with \autoref{lem_balanced} in mind: balanced wirings behave well with respect to height of terms.

\begin{definition}[computation space]\label{def-comp}
	Given a balanced wiring $\wiring{F}$, we define its \emph{computation space}
	$\comp{\wiring{F}}$ as the set of facts of height at most $\h(\wiring{F})$,
	built using only the symbols appearing in $\wiring{F}$ and the constant symbol $\factsym$.
\end{definition}

\begin{lemma}[separation]\label{lem_sep}
	If $F$ is balanced, then $\comp{\wiring{F}}$ is separating for~$F$.
\end{lemma}

\begin{proof}
	By \autoref{lem_balanced}, $\wiring F(\fact u)$ is of height at most $\max\{\h(\wiring F),\h(u)\}\leq\h(\wiring F)$ and it contains only symbols occurring in $\wiring F$ and $u$,
	therefore if $\BF u\in \comp{\wiring{F}}$ we have $\wiring F\BF u\subseteq \comp{\wiring{F}}$.
	
	By \autoref{lem_balanced} again, $F^n$ is still of height at most $\h(F)$. If $(F^n)\BF u=0$ for all $\BF u\in\comp{\wiring{F}}$, it means the flows of $F^n$ do not match any closed term of height at most $\h(F)$ built with the symbols occurring in $F$ (and eventually $\factsym$). This is only possible if $F$ contains no flow, \textit{ie.} $F=0$.\qed
\end{proof}

As $F$ is a finite set, thus built with finitely many symbols, $\comp{\wiring{F}}$ is also a finite set.
We can be a little more precise and give a bound to its cardinal.

\begin{proposition}[cardinality]\label{card-comp}
	Let $\wiring{F}$ be a balanced wiring, $A$ the maximal arity of function symbols
	occurring in $\wiring{F}$ and $S$ the set of symbols occurring in $\wiring{F}$,
	then $\card{\comp{\wiring{F}}}\ \leq\ (\card S+1)^{P_{\h(\wiring{F})}(A)}$,
	where $P_k(X)=1+X+\cdots+X^k$.
\end{proposition}

\begin{proof}
	The number of terms of height $\h(F)$ built over the set of symbols $S\cup\{\factsym\}$
	of arity bounded by $A$ is at most as large as the number of complete trees
	of degree $A$ and height $\h(F)$ 
	(that is, trees where nodes of height less than $\h(F)$ have exactly $A$ childs),
	with nodes labeled by elements of $S\cup\{\factsym\}$.\qed
\end{proof}

Then, we can encode in a directed graph\footnote{Here by directed graph we mean a set of \emph{vertices} $V$ together with a set of \emph{edges} $E\subseteq V\times V$.
We say that there is an edge \emph{from $e\in V$ to $f\in V$} when $(e,f)\in E$.} the action of the wiring on its computation space.

\begin{definition}[computation graph]\label{compgraph}
	If $F$ is a balanced wiring, we define its \emph{computation graph} $\G(\wiring{F})$ as the directed graph:
	\begin{itemize}
		\item The vertices of $\G(\wiring{F})$ are the elements of $\comp{\wiring{F}}$.
		\item There is an edge from $\BF u$ to $\BF v$ in $\G(\wiring{F})$ if $\BF v\in \wiring{F} \BF u$.
	\end{itemize}
\end{definition}

We state finally that the computation graph of a wiring contains enough information on the latter to determine its nilpotency.
This is a key ingredient in the proof of \autoref{nilp-in-l}, as the research of paths and cycles in graphs are problems that are well-known to be solvable within logarithmic space.

\begin{lemma}\label{nilp-iff-acycl}
	A balanced wiring $\wiring{F}$ is nilpotent (\autoref{def_nilp}) \iff $\G(\wiring{F})$ is acyclic.
\end{lemma}

\begin{proof}
	Suppose there is a cycle of length $n$ in $\G(\wiring{F})$, and let $\BF u$ be the label of a vertex which is part of this cycle. By definition of $\G(\wiring{F})$, $\BF u\in(\wiring{F}^n)^k\BF u$ for all $k$, which means that $(\wiring{F}^n)^k\neq0$ for all $k$ and therefore $F$ cannot be nilpotent.
	
	Conversely, suppose there is no cycle in $\G(\wiring{F})$. As it is a finite graph, this entails a maximal length $N$ of paths in $\G(F)$. By definition of $\G(\wiring{F})$, this means that $F^{N+1}\BF u=0$ for all $\BF u\in\comp{\wiring{F}}$ and with \autoref{lem_sep} we get $F^{N+1}=0$. \qed
\end{proof}

Moreover, the computation graph of a deterministic (\autoref{def_fun}) wiring has a specific shape, which in turn induces a deterministic procedure in this case.

\begin{lemma}\label{det-graph}
	If $F$ is a balanced and deterministic wiring, $\G(\wiring{F})$ has an out-degree (the maximal number of edges a vertex can be the source of) bounded by $1$.
\end{lemma}

\begin{proof}
	It is a direct consequence of the definitions of $\G(\wiring{F})$ and determinism.
	\qed
\end{proof}

	\subsection{Tensor product and other semirings}
	Finally, we list a few other semirings that will be used in the next section, where we define the notions of representation of a word and observation.

\smallskip
The binary function symbol $\p$ can be used to define an operation that is similar to the algebraic notion of tensor product.

\begin{definition}[tensor product]\label{ptensor}
	Let $\,u\flow v\,$ and $\,t\flow w\,$ be two flows.
	Suppose we have chosen representatives of their renaming classes that have disjoint sets of variables. We define their \emph{tensor product} as
	$(u\flow v) \ptensor (t\flow w):=\: u\p t \flow v\p w$.
	The operation is extended to wirings by $(\sum_i f_i)\ptensor(\sum_j g_j):= \sum_{i,j} f_i\ptensor g_j$.
%
	Given two semirings $\C A,\C B$, we define 
%
	$\C A\ptensor \C B:= \,\set{\sum_i F_i\ptensor G_i}{F_i\in\C A\,,\,G_i\in \C B}$.
\end{definition}

The tensor product of two semirings is easily shown to be a semiring.

\example{~

$(\TT f(x)\p y\flow y\p x)\ptensor(x\flow \TT g(x))= (\TT f(x)\p y)\p z\flow (y\p x)\p \TT g(z) $}

\begin{notation}
	As the symbol $\p$, the $\ptensor$ operation is not associative. We carry on the convention for $\p$ and write it as \emph{right associating}: $\C A\ptensor\C B\ptensor\C C:=\C A\ptensor(\C B\ptensor\C C)$.
\end{notation}

Semirings can also naturally be associated to any set of closed terms or to the restriction to a certain set of symbols.

\begin{definition}
	Given a set of closed terms $E$, we define the following semiring $\calg E:=\set{\sum_i t_i\flow u_i}{t_i,u_i \in E}$.
	If $\TT S$ is a set of symbols and $\C A$ a semiring, we write $\ralg{\TT S}{\C A}$ the semiring of wirings of $\C A$ , that do not use the symbols in $\TT S$.
\end{definition}

This operation yields semirings because composition of flows made of closed terms involves no actual unification: it is just equality of terms and therefore one never steps out of $\calg E$.

\smallskip
Finally, the unit $I=x\flow x$ of $\ualg$ yields a semiring.

\begin{definition}[unit semiring]
	We call \emph{unit semiring} the semiring $\unitalg:=\{\,\unit\,\}$.
\end{definition}

\section{Words and Observations}\label{sec_representation}
We define in this section the global framework that will be used later on to obtain the characterization of logarithmic space computation.
In order to discuss the contents of this section, let us first define two specific semirings.

\begin{definition}[word and observation semirings]
	We fix two (disjoint) infinite sets of constant symbols $\TT P$ and $\TT S$, and a unary function symbol $\M$.
	We denote by $\M(\TT P)$ the set of terms $\M(\TT p)$ with $\TT p\in\TT P$.
	We define the following two semirings that are included in $\balg$:
	\begin{itemize}	
	\item The \emph{word semiring} is the semiring $\walg:=\C \unitalg\ptensor\unitalg\ptensor \calg{\TT M(\TT P)}$.
	\item The \emph{observation semiring} is the semiring $\oalg:=\calg{\TT S}\ptensor\ralg{\TT P}{\balg}$.
	\end{itemize}
\end{definition}

These two semirings will be used as parameters of a construction $\matr{\Sigma}{.}$ to define the representation of words and a notion of abstract machine, that we shall call observations, on an alphabet $\Sigma$ (we suppose $\star\not\in\Sigma$). 

\begin{definition}
	We fix the set of constant symbols $\LR:=\{\L,\R\}$.
	
	Given a set of constant symbols $\Sigma$ and a semiring $\C A$ we define the semiring $\matr{\Sigma}{\C A}:=\calg{(\Sigma\cup\{\star\})} \ptensor \calg\LR\ptensor\C A$.
\end{definition}

In the following of this section, we will show how to represent lists of elements of $\Sigma$ by 
wirings in the semiring $\matr{\Sigma}{\walg}$. 
Then, we will explain how the semiring $\matr{\Sigma}{\oalg}$ 
captures a notion of abstract machine.
In the last section of the paper we will explain further how observations and words interact, and prove that this interaction captures logarithmic space computation.

	\subsection{Representation of Words}\label{sec_words}
	We now show how one can represent words by wirings in $\matr{\Sigma}{\walg}$. We recall this semiring is defined as 
\(\big(\calg{(\Sigma\cup\{\star\})} \ptensor \calg\LR\big)\ptensor\C \unitalg\ptensor\unitalg\ptensor \calg{\TT M(\TT P)}\).

The part $\calg{(\Sigma\cup\{\star\})} \ptensor \calg\LR$ deals with, and is dependent on, the alphabet $\Sigma$ considered;
this is where the integer and the observation will interact.
The two instances of the unit semiring $\unitalg$ correspond to the fact that the word cannot affect parts of the observation
that correspond to internal configurations.
The last part, namely the semiring $\calg{\TT M(\TT P)}$, will contain the \emph{position constants} of the representation of words.



\begin{notation}
	We write $t\sflow u$ for $t\flow u +u\flow t$.
\end{notation}

\begin{definition}[word representations]\label{word-rep}
	Let $W= \TT c_1\dots\TT c_n$ be a word over an alphabet $\Sigma$ and $p=\extlist{\TT p_0,\TT p_1}{\TT p_n}$ be pairwise distinct constant symbols.
	
	Writing $\TT p_{n+1}=\TT p_{0}$ and $\TT{c_{0}} = \TT{c_{n+1}} = \start$, we define the \emph{representation} of $W$ associated with $\extlist{\TT p_0,\TT p_1}{\TT p_n}$ as the following wiring: 
	\begin{equation}\label{rep-entier-sum}
	\bar{W}_{p}= \sum_{i=0}^{n}{\wordterm{c_{i}}{\L}{x}{y}{\TT p_{i}} \sflow {\wordterm{c_{i+1}}{\R}{x}{y}{\TT p_{i+1}}}}
	\end{equation}

	We will write $\setofrep{W}$ the set of representations of a given word $W$.
\end{definition}

To understand better this representation, consider that each symbol in the alphabet $\Sigma$ comes in 
two \enquote{flavors}, \emph{left} and \emph{right}. 
Then, one can easily construct the \enquote{context} 
\(\bar{W}=\sum_{i=0}^{n}{\wordterm{\TT c_{i}}{\L}{x}{y}{[~]_{i}} \sflow {\wordterm{\TT c_{i+1}}{\R}{x}{y}{[~]_{i+1}}}}\)
from the list as the sums of the arrows in the following picture (where $x$ and $y$ are omitted): 
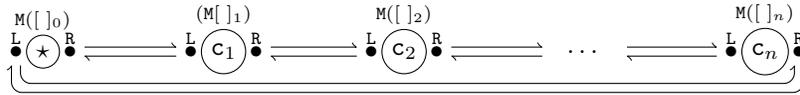
\begin{center}
\begin{tikzpicture}[x=1.2cm,y=1.2cm,inner sep=2pt, outer sep=1pt]
	\node[shape=circle,draw] (S) at (0,0) {$\star$};
		\node (SE) at (S.east) {};
		\node (SW) at (S.west) {};
		\node (SN) at (S.north) {};
		\node (So) at (SE.east) {$\bullet$};
		\node (SoL) at (So.north) {\scriptsize{$\R$}};
		\node (Si) at (SW.west) {$\bullet$};
		\node (SiL) at (Si.north) {\scriptsize{$\L$}};
		\node (Sv) at ($(SN.north)+(0,0.05)$) {\scriptsize{$\TT M([~]_{0})$}};
	\node[shape=circle,draw] (A) at (2,0) {$\TT c_{1}$};
		\node (AE) at (A.east) {};
		\node (AW) at (A.west) {};
		\node (AN) at (A.north) {};
		\node (Ao) at (AE.east) {$\bullet$};
		\node (AoL) at (Ao.north) {\scriptsize{$\R$}};
		\node (Ai) at (AW.west) {$\bullet$};
		\node (AiL) at (Ai.north) {\scriptsize{$\L$}};
		\node (Av) at ($(AN.north)+(0,0.05)$) {\scriptsize{$(\TT M[~]_{1})$}};
	\node[shape=circle,draw] (B) at (4,0) {$\TT c_{2}$};
		\node (BE) at (B.east) {};
		\node (BW) at (B.west) {};
		\node (BN) at (B.north) {};
		\node (Bo) at (BE.east) {$\bullet$};
		\node (BoL) at (Bo.north) {\scriptsize{$\R$}};
		\node (Bi) at (BW.west) {$\bullet$};
		\node (BiL) at (Bi.north) {\scriptsize{$\L$}};
		\node (Bv) at ($(BN.north)+(0,0.05)$) {\scriptsize{$\TT M([~]_{2})$}};
	\node[shape=circle] (C) at (6,0) {$\dots$};
		\node (CE) at (C.east) {};
		\node (CW) at (C.west) {};
		\node (Co) at (CE.east) {};
		\node (Ci) at (CW.west) {};
	\node[shape=circle,draw] (D) at (8,0) {$\TT c_{n}$};
		\node (DE) at (D.east) {};
		\node (DW) at (D.west) {};
		\node (DN) at (D.north) {};
		\node (Do) at (DE.east) {$\bullet$};
		\node (DoL) at (Do.north) {\scriptsize{$\R$}};
		\node (Di) at (DW.west) {$\bullet$};
		\node (DiL) at (Di.north) {\scriptsize{$\L$}};
		\node (Dv) at ($(DN.north)+(0,0.05)$) {\scriptsize{$\TT M([~]_{n})$}};
	\draw[-left to] ($(So.east)+(0,0.05)$) to ($(Ai.west)+(0,0.05)$) {};
	\draw[-left to] ($(Ai.west)+(0,-0.05)$) to ($(So.east)+(0,-0.05)$) {};
	\draw[-left to] ($(Ao.east)+(0,0.05)$) to ($(Bi.west)+(0,0.05)$) {};
	\draw[-left to] ($(Bi.west)+(0,-0.05)$) to ($(Ao.east)+(0,-0.05)$) {};
	\draw[-left to] ($(Bo.east)+(0,0.05)$) to ($(Ci.west)+(0,0.05)$) {};
	\draw[-left to] ($(Ci.west)+(0,-0.05)$) to ($(Bo.east)+(0,-0.05)$) {};
	\draw[-left to] ($(Co.east)+(0,0.05)$) to ($(Di.west)+(0,0.05)$) {};
	\draw[-left to] ($(Di.west)+(0,-0.05)$) to ($(Co.east)+(0,-0.05)$) {};
	\draw[-left to, rounded corners] ($(Do.south)+(0.05,0)$)  -- ($(Do.south)+(0.05,-0.3)$) -- ($(Si.south)+(-0.05,-0.3)$) -- ($(Si.south)+(-0.05,0)$);
	\draw[-left to, rounded corners] ($(Si.south)+(0.05,0)$)  -- ($(Si.south)+(0.05,-0.2)$) -- ($(Do.south)+(-0.05,-0.2)$) -- ($(Do.south)+(-0.05,0)$);
\end{tikzpicture}
\end{center}
Then, choosing a set $p=\extlist{\TT p_{0}}{\TT p_{n}}$ of position constants, intuitively representing physical memory addresses,
the representation $\bar{W}_{p}$ of a word associated with $p$ is obtained by filling, for all $i=\extlist{0}{n}$, the hole $[~]_{i}$ by the constant $\TT p_{i}$.

\smallskip
This abstract representation of words is not an arbitrary choice.
It comes from the interpretation of lists in geometry of interaction.

Indeed, in \systF, the type of binary lists corresponds to the formula 
$\forall X~(X\Rightarrow X)\Rightarrow (X\Rightarrow X)\Rightarrow (X\Rightarrow X)$.
Any lambda-term in normal form of this type can be written as
$\lambda f_{0}f_{1}x.\,f_{\TT c_{1}}f_{\TT c_{2}}\cdots f_{\TT c_{k}}x$,
where $\TT c_{1}\cdots c_k$ is a word on $\{0,1\}$.
The \GoI representation of such a lambda-term yields the abstract representation just defined\footnote{A thorough explanation can be found in previous work by Aubert and Seiller~\cite{Aubert2012}.}.
Notice that the additional symbol $\star$ used to represent words corresponds to the variable $x$ in the preceding lambda-term.
Note also the fact that the representation of integer is \emph{cyclic}, and that the symbol $\star$ serves as a reference for the starting/ending point of the word.

Let us finally stress that the words are represented as \emph{deterministic} wirings.
This implies that the restriction to deterministic observations will correspond to restricting ourselves to deterministic pointer machines.
The framework, however, allows for a number of generalization and variants.
For instance, one can define a representation of trees by adapting \autoref{word-rep} in such a way that every vertex is related to its descendants; doing so would however yield non-deterministic wirings.
In the same spirit, a notion of \enquote{one-way representations of words}, defined by replacing the symbol $\sflow$ by the symbol $\flow$ in \eqref{rep-entier-sum} of \autoref{word-rep}, could be used to characterize one-way multi-heads automata.


	\subsection{Observations}\label{sec_observations}
	We now define \emph{observations}.
We will then explain how these can be thought of as a kind of abstract machines.
An observation is an element of the semiring
\[\calg{(\Sigma\cup\{\star\})} \ptensor \calg\LR\ptensor(\calg{\TT S}\ptensor\ralg{\TT P}{\balg})\]
Once again, the part of the semiring $\calg{(\Sigma\cup\{\star\})} \ptensor \calg\LR$ is dependent on the alphabet $\Sigma$ considered and represents the point of interaction between the words and the machine. 
The semiring $\calg{\TT S}$ intuitively corresponds to the \emph{states} of the observation, 
while the part $\ralg{\TT P}{\balg}$ forbids the machine to act non-trivially on the \emph{position constant} of the representation of words. 
The fact that the machine cannot perform any operation on the memory addresses --~the position constants~-- of the word representation explains why observations are naturally though of as a kind of \emph{pointer machines}.

\begin{definition}[observation]
	An \emph{observation} is any element $\obs{O}$ of $\matr{\Sigma}{\oalg}$.
\end{definition}

We can define the language associated to an observation.
The condition of acceptance will be represented as the nilpotency of the product $\obs{O}\bar{W}_{p}$ where $\bar{W}_{p}\in\setofrep{W}$ represents a word $W$ and $\obs{O}$ is an observation.


\begin{definition}[language of an observation]
\label{lang-obs}
Let $\obs{O}$ be an observation on the alphabet $\Sigma$.
We define the \emph{language accepted by $\obs{O}$} as

\smallskip
\centerline{
\(\lang{\obs{O}}:=\set{W\in \Sigma^{\ast}}{\forall p, \,\isnilp{\obs{O}\bar{W}_{p}}}\)
}

\end{definition}

One important point is that the semirings $\matr{\Sigma}{\walg}$ and $\matr{\Sigma}{\oalg}$ are not completely disjoint, and therefore allow for non-trivial interaction of observations and words. 
However, they are sufficiently disjoint so that this computation does not depend on the choice of the representative of a given word. 

\begin{lemma}
\label{equiv-of-rep}
Let $W$ be a word, and $\bar{W}_{p}, \bar{W}_{q}\in\setofrep{W}$.
For every observation $\obs{O}\in \matr{\Sigma}{\oalg}$, $\isnilp{\obs{O}\bar{W}_{p}}$ if and only if $\isnilp{\obs{O}\bar{W}_{q}}$.
\end{lemma}

\begin{proof}
As we pointed out, the observation cannot act on the position constants of the representations $\bar{W}_{p}$ and $\bar{W}_{q}$.
This implies that for all integer $k$ the wirings $(\obs{O}\bar{W}_{p})^{k}$ and $(\obs{O}\bar{W}_{q})^{k}$ are two instances of the same \emph{context}, \ie they are equal up to the interchange of the positions constants $\extlist{\TT p_{0}}{\TT p_{n}}$ and $\extlist{\TT q_{0}}{\TT q_{n}}$.
This implies that $(\obs{O}\bar{W}_{p})^{k}=0$ if and only if $(\obs{O}\bar{W}_{q})^{k}=0$.
\qed
\end{proof}


\begin{corollary}%
\label{lang-reloaded}
Let $\obs{O}$ be an observation on the alphabet $\Sigma$.
The set $\lang{\obs{O}}$ can be equivalently defined as the set

\smallskip
\centerline{
\(\lang{\obs{O}}=\set{W\in \Sigma^{\ast}}{\exists p,\, \isnilp{\obs{O}\bar{W}_{p}}}\)
}
\end{corollary}

This result implies that the notion of acceptance has the intended sense and is finitely verifyable: whether a word $W$ is accepted by an observation $O$ can be checked
without considering all representations of $W$.


This kind of situation where two semirings $\walg$ and $\oalg$ are disjoint enough to obtain \autoref{lang-reloaded} can be formalized through the notion of \emph{normative pair} considered in earlier works \cite{Girard2012,Aubert2012,Aubert2014}.

%

\section{Logarithmic Space}\label{sec_complexity}
	This section starts by explaining the computation one can perform with the observations, and prove that it corresponds to logarithmic space computation by showing how pointer machines can be simulated.
Then, we will prove how the language of an observation can be decided within logarithmic space.

This section uses the complexity classes \logspace and \conlogspace, as well as notions of completeness of a problem and reduction between problems.
We use in \autoref{sec_soundness} the classical theorem of \conlogspace-completeness of the acyclicity problem in directed graphs,
 and in \autoref{sec_comleteness} a convenient model of computation, two-ways multi-heads finite automata, a generalization of automata also called \enquote{pointer machine}.
The reader who would like some reminders may have a look at any classical textbook~\cite{Savage1998}.
Note that the non-deterministic part of our results concerns \conlogspace, or equivalently \nlogspace by the famous Immerman-Szelepcsényi theorem.

	\subsection{Completeness: Observations as Pointer Machines}\label{sec_comleteness}
	This section describes the kind of computation one can model with observations.

Let $h_0, x, y$ be variables, $\TT p_0, \TT p_1, \nary 0$ constants and $\Sigma = \{0, 1\}$, the excerpt of a dialogue in \autoref{fig-dial} between an observation $O=o_1 + o_2 + \cdots\,$ and the representation of a word $\bar W_p=w_1 + w_2 + \cdots\,$ should help the reader to grasp the mechanism.

\begin{figure}
\begin{align}
\wordterm{\start}{\L}{\state{init}}{\nary 0}{h_0}	& \flow \wordterm{\start}{\R}{\state{init}}{\nary 0}{h_0}	\tag{$o_1$}\\
\wordterm{\start}{\R}{x}{y}{\TT p_0 } & \flow \wordterm{1}{\L}{x}{y}{\TT p_1 } \tag{$w_1$}\\
\wordterm{1}{\L}{\state{init}}{\nary 0}{h_0} & \flow \wordterm{1}{\L}{\state{b}}{\nary 0}{h_0}	\tag{$o_2$}\\
\wordterm{1}{\L}{x}{y}{\TT p_1} &  \flow \wordterm{\start}{\R}{x}{y}{\TT p_0 } \tag{$w_2$}\\
\shortintertext{By unification,}
\wordterm{\start}{\L}{\state{init}}{\nary 0}{\TT p_0} & \flow \wordterm{1}{\L}{\state{init}}{\nary 0}{\TT p_1 }	\tag{$o_1 w_1$}\\
\wordterm{\start}{\L}{\state{init}}{\nary 0}{\TT p_0} & \flow \wordterm{1}{\L}{\state{b}}{\nary 0}{\TT p_1}	\tag{$o_1 w_1 o_2$}\\
\wordterm{\start}{\L}{\state{init}}{\nary 0}{\TT p_0} & \flow \wordterm{\start}{\R}{\state{b}}{\nary 0}{\TT p_0}	\tag{$o_1 w_1 o_2 w_2$}
\end{align}
This can be understood as the small following dialogue:
\begin{dialogue}
	\speak{$o_1$} \direct{Is in state $\state{init}$} \enquote{I read $\start$ from left to right, what do I read now?}
	\speak{$w_1$} \enquote{Your position was $\TT p_0$, you are now in position $\TT p_1$ and read $1$.}
	\speak{$o_2$} \direct{Change state to $\state{b}$} \enquote{I do an about-turn, what do I read now?}
	\speak{$w_2$} \enquote{You are now in position $\TT p_0$ and read $\start$.}
\end{dialogue}
\caption{An example of dialogue between an observation and the representation of a word}
\label{fig-dial}
\end{figure}

We just depicted two transitions corresponding to an automata that reads the first bit of the word, and if this bit is a $1$, goes back to the starting position, in state $\state{b}$.
Remark that the answer of $w_1$ differs from the one of $w_2$: there is no need to clarify the position (the variable argument of $\M$), since $h_0$ was already replaced by $\TT p_1$.
Such an information is needed only in the first step of computation: after that, the updates of the position of the pointer take place on the word side.
Remark that neither the state nor the constant $\nary 0$ is an object of dialogue.

Note also that this excerpt corresponds to a deterministic computation. In general, several elements of the observation could get unified with the current configuration, yielding non-deterministic transitions.

\subsubsection{Multiple Pointers and Swapping}
We now add some computational power to our observation by adding the possibility to handle several pointers.
The observations will now use a $k$-ary function $\nary k$ that allows to \enquote{store} $k$ positions.
This part of the observation is not affected by the word, which means that only one head (the \emph{main pointer}) can move.
The observation can exchange the position of the main pointer and the position stored in $\nary k$: we therefore refer to the arguments of $\nary k$ as \emph{auxiliary pointers} that can become the main pointer at some point of the computation.
This is of course strictly equivalent to several heads that have the ability to move.

Consider the following flow, that encodes the transition \enquote{if the observation reads $1\p\R$ in state $\state{s}$, it stores the position of the main pointer (the variable $h_0$) at the $i$-th position in $\nary k$ and start reading the input with a new pointer}:

\medskip
\centerline{
\(
\wordterm{1}{\R}{\state{s}}{\nary k (\extlist{h_1}{\extlist{h_i}{h_k}})}{h_0}  \flow \wordterm{\start}{\R}{\state{s'}}{\nary k (\extlist{h_1}{\extlist{h_0}{h_k}})}{h_i}
\)
}

\medskip
Suppose now we want to write an observation that, if it reads $0\p\L$ in state $\state{r}$, then it restores the value that was stored in the $i$-th position of $\nary k$, it stores the position in $\M$ and changes to state $\state{r'}$, how should we proceed?
The point is that the flow above did not \enquote{memorize} the value that was stored at that position.

\medskip
\centerline{
\(
\wordterm{0}{\L}{\state{r}}{\nary k (\extlist{h_1}{\extlist{h_0}{h_k}})}{h_i} \flow \wordterm{x}{\_}{\state{r'}}{\nary k (\extlist{h_1}{\extlist{h_i}{h_k}})}{h_0}
\)
}

\medskip
The variable $x$ should be free and will be unified with the flow of the word that corresponds to the position constant that was substituted to $h_0$.
But there are two such flows in the word, so the direction of the next movement should be indicated in the $\_$ slot.
Such an information could be encoded in the state $\state{r'}$%
\footnote{That is, we could have two states $\state{r'_{\L}}$ and $\state{r'_{\R}}$ and two flows accordingly.}
and in the transition function.

\subsubsection{Acceptance and Rejection}
Remember (\autoref{lang-reloaded}) that the language of an observation is the set of words such that the wiring composed of the observation applied to a representation of the word is nilpotent.
So one could add a flow with the head corresponding to the desired situation leading to acceptance, and the body being $0$.
But in fact, it is sufficient not to add any flow: doing nothing is accepting!

The real challenge is to reject a word: it means to loop forever.
We cannot simply add the unit ($\unit := x \flow x$) to our observation, since that would make our observation loop \emph{no matter the word in input}.
So we have to be clever than that, and to encode rejection as a re-initialization of the observation: we want the observation to put all the pointers on $\start$ and to go back to an $\state{init}$ state.
So, a loop is in fact a \enquote{perform for ever the same computation}.

If we reject, suppose the main pointer was reading from right to left, and that we changed to $\state{b}$.
Then, for every $\TT c \in \Sigma$, it is enough to add the transitions \eqref{reject} and \eqref{reinit} to the observation,

\vspace{-6mm}
\begin{align}
\wordterm{\TT c}{\R}{\state{b}}{\A(\extlist{h_1}{h_k})}{h_0}	& \flow \wordterm{\TT c}{\L}{\state{b}}{\A(\extlist{h_1}{h_k})}{h_0} \label{reject} \tag{go-back-$\TT c$}\\
\wordterm{\start}{\R}{\state{b}}{\A(\extlist{h_1}{h_k})}{h_0}	& \flow \wordterm{\start}{\R}{\state{init}}{\A(\extlist{h_0}{h_0})}{h_0} \tag{re-init} \label{reinit}
\end{align}
\vspace{-6mm}

Once the main pointer is back on $\start$, \eqref{reinit} re-initializes all the positions of the auxiliary pointers to the position of $\start$ and changes the state for $\state{init}$.

There is another justification for this design: as the observation and the representation of the word are sums, and as the computation is the application, any transition that can be applied will be applied, \ie if the body (the right-member) of a flow of our observation and the head (the left-member) of a flow of the word can be unified, the computation will start in a possibly \enquote{wrong} initialization.
That some of this incorrect runs accept for incorrect reason is no trouble, since only rejection is \enquote{meaningful} due to the nilpotency criterion.
But, with this framework, an incorrect run will be re-initialized to the \enquote{right} initialization, and performs the correct computation: in that case, it will loop if and only if the input is rejected.

\subsubsection{Two-Ways Multi-Heads Finite Automata and Completeness}
The model we just developed has clearly the same expressivity as two-ways multi-heads finite automata, a model of particular interest to us for it is well studied, tolerant to a lot of enhancements or restrictions%
\footnote{In fact, most of the variations (the automata can be one-way, sweeping, rotating, oblivious, etc.) are studied in terms of number of states needed to simulate a variation with another, but most of the time they characterize the same complexity classes.
}
and gives an elegant characterization of \logspace and \nlogspace~\cite{Holzer2008,Pighizzini2013}.

Then, by a plain and uniform encoding of two-ways multi-heads finite automata, we get \autoref{th_compl}.
That acceptance and rejection in the non-deterministic case are \enquote{reversed} (\ie all path have to accept for the computation to accepts) makes us characterize \conlogspace instead of \nlogspace.

Note that encoding a \emph{deterministic} automaton yields a wiring of the form of \autoref{lem_det}, which would be therefore a deterministic wiring.
\begin{theorem}\label{th_compl}
If $L\in\conlogspace$, then there is an observation $O$ such that $\lang O=L$. Moreover, if $L\in \logspace$, $O$ can be chosen deterministic.
\end{theorem}

	\subsection{Soundness of Observations}\label{sec_soundness}
	We now use the results of \autoref{sec_compgraph} and \autoref{sec_observations} to design a procedure that decides whether a word belongs to the language of an observation within logarithmic space.
This procedure will use the classical reduction from a problem to testing the acyclicity of a graph, a problem well-known to be tractable with logarithmic space resources.

First, we show how the computation graph of the product of the observation and the word representation can be constructed deterministically using only logarithmic space;
then, we prove that testing the acyclicity of such a graph can be done within the same bounds.
Here, depending on the shape of the graph (which is dependent in itself of determinism of the wiring, 
recall \autoref{det-graph}), the procedure will be deterministic or non-deterministic.

Finally, using classical composition of logspace algorithms%
\footnote{The reader unused to the mechanism of composition of logspace-algorithms may have a look at a classical textbook \cite[Fig. 8.10]{Savage1998}.}, \autoref{nilp-iff-acycl} and \autoref{lang-reloaded}, we will obtain the result expected, that is:

\begin{theorem}%
\label{nilp-in-l}
If $O$ is an observation, then $\lang O\in \conlogspace$. If moreover $O$ is deterministic, then $\lang O\in\logspace$.
\end{theorem}

\subsubsection{A Foreword on Word and Size}
Given a word $W$ over $\Sigma$, to build a representation $\bar{W}_p$ as in \autoref{word-rep} is clearly in $\logspace$: it is a plain matter of encoding.
By \autoref{equiv-of-rep}, it is sufficient to consider a single representation.
So for the rest of this procedure, we consider given $\bar{W}_{p} \in \setofrep{W}$ and write $\wiring{F}:= \obs{O}\bar{W}_{p}$.
The size of $\Sigma$ is a constant, and it is clear that the maximal arity and the height of the balanced wiring $\wiring{F}$ remain fixed when $W$ varies.
The only point that fluctuates is the cardinality of the set of symbols that occurs in $\wiring{F}$, and it is linearly growing with the length of $W$, corresponding to the number of positions constant.
In the following, any mention to a logarithmic amount of space is to be completed by \enquote{relatively to the length of $W$}.

\subsubsection{Building the Computation Graph}
We need two main ingredients to build the computation graph (\autoref{compgraph}) of $\wiring{F}$: to enumerate the computation space $\comp{\wiring{F}}$ (recall \autoref{def-comp}), and to determine whether there is an edge between two vertices.

By \autoref{card-comp}, $\card{\comp{\wiring{F}}}$ is polynomial in the size of $W$.
Hence, given a balanced wiring $\wiring{F}$, a logarithmic amount of memory is enough to enumerate the members of $\comp{\wiring{F}}$, that is the vertices of $\G(\wiring{F})$.

Now the second part of the construction of $\G(\wiring{F})$ is to determine if there is an edge between two vertices.
Remember that there is an edge from $\BF u = \fact{u}$ to $\BF v = \fact{v}$ in $\G (\wiring{F})$ if $\BF v \in \wiring{F} \BF u$.
So one has to scan the members of $\wiring{F}=\obs{O}\bar{W}_{p}$: if there exists $(t_1 \flow t_2)(t'_1 \flow t'_2) \in \wiring{F}$ such that $(t_1 \flow t_2)(t'_1 \flow t'_2) (\fact{u}) = \fact{v}$, then there is an edge from $\BF u$ to $\BF v$.
To list the members of $\wiring{F}$ is in $\logspace$, but unification in general is a difficult problem (see \autoref{ssec_unif}).
The special case of matching can be tested with a logarithmic amount of space:

\begin{theorem}[Matching is in {\sc \logspace} \protect{\cite[p.~49]{Dwork1984}}]
\label{thm-unif-logspace}
	Given two terms $t$ and $u$ such that either $t$ or $u$ is closed, deciding if they are matchable is in \logspace.
\end{theorem}

Actually, this result relies on a subtle manipulation of the representation of the terms as \emph{simple directed acyclic graphs}\footnote{The reader willing to learn more about this may have a look at a classical chapter~\cite{Baader2001}.}, where the variables are \enquote{shared}.
Translations between this representation of terms and the usual one can be performed in logarithmic space \cite[p.~38]{Dwork1984}.

\subsubsection{Deciding if $\G(\wiring{F})$ is Acyclic}
We know thanks to \autoref{nilp-iff-acycl} that answering this question is equivalent to deciding if $\wiring{F}$ is nilpotent.
We may notice that $\G(\wiring{F})$ is a directed, potentially unconnected graph of size $\card{\comp{\wiring{F}}}$.

It is well-know that testing for acyclicity of a directed
graph is a \conlogspace \cite[p.~83]{Jones1975} problem.
Moreover, if $\wiring{F}$ is deterministic (which is the case when $O$ is), then $\G(\wiring{F})$ has out-degree bounded by $1$ (\autoref{det-graph}) and one can test its acyclicity without being non-deterministic: it is enough to list the vertices of $\comp{\wiring{F}}$, and for each of them to follow $\card{\comp{\wiring{F}}}$ edges and to test for equality with the vertex picked at the beginning.
If a loop is found, the algorithm rejects, otherwise it accepts after testing the last vertex.
Only the starting vertex and the current vertex need to be stored, which fits within logarithmic space, and there is no need to do any non-deterministic transitions.

\section{Conclusion}
	We presented the unification semiring, a construction that can be used both as an algebraic model of logic programming and as a setting for a dynamic model of logic.
Within this semiring, we were able to identify a class of wirings that have the exact expressive power of logarithmic space computation.

\smallskip

If we try to step back a little, we can notice that the main tool in the soundness proof (\autoref{sec_soundness}) is the computation graph, defined in \autoref{sec_compgraph}.
More precisely, the properties of this graph, notably its cardinal (that turns out to be polynomial in the size of the input), allow to define a decision procedure that needs only logarithmic space.
The technique is modular, hence not limited to logarithmic space: identifying other conditions on wirings that ensure size bounds on the computation graph would be a first step towards the characterization of other space complexity classes.

\smallskip
Concerning completeness, the choice of encoding pointer machines (\autoref{sec_comleteness}) rather than log-space bounded Turing machines was quite natural.
Balanced wirings correspond to the idea of computing with pointers: manipulation of data without writing abilities, and thus with no capacity to store any information other than a fixed number of  positions on the input.

By considering other classes of wirings or by modifying the encoding it might be possible to capture other notions of machines characterizing some complexity classes: we already mentioned that a modifcation of the representation of the input that would model one-way finite automata.

\smallskip
The relation with proof theory needs to be explored further: the approach of this paper seems indeed to suggest a sort of \enquote{Curry-Howard} correspondance for logic programming.

As \autoref{ssec_goilp} highlighted, there are many notions that might be transferable from one field to the other,
thanks to a common setting provided by geometry of interaction and the unification semiring.
Most notably, the notion of nilpotency (on the proof-theoretic side: strong normalization) corresponds to a variant of boundedness of logic programs, a property that is usually hard to ensure.

Another direction could be to look for a proof-system counterpart of this work: a corresponding \enquote{balanced} logic of logarithmic space.






\bibliographystyle{splncs03}
\bibliography{court,biblio}
\addcontentsline{toc}{section}{References}

\end{document}